\documentclass[english, aps, prl, preprint, superscriptaddress, colorlinks=true, citecolor=blue, linkcolor=red, urlcolor=blue]{revtex4-2}

\usepackage{braket}
\usepackage{fancyhdr,graphicx}
\usepackage[table,xcdraw]{xcolor}
\usepackage[normalem]{ulem}
\useunder{\uline}{\ul}{}
\usepackage[T1]{fontenc}
\usepackage{chemformula} 
\usepackage[utf8x]{inputenc}
\usepackage{gensymb}
\usepackage{times}
\usepackage{color}
\usepackage{amsfonts}
\usepackage{amssymb}
\usepackage[colorlinks,bookmarks=false,citecolor=blue,linkcolor=red,urlcolor=blue]{hyperref}
\input pdfcolor.tex
\usepackage{colortbl,amsthm,amsmath,amssymb,txfonts}
\usepackage{graphicx}
\usepackage{epsfig,color}
\usepackage{verbatim}
\usepackage{dcolumn}
\usepackage[normalem]{ulem}
\usepackage{bm}
\usepackage{xcolor}

\usepackage{orcidlink}
\usepackage[normalem]{ulem}
\usepackage{cancel}
\usepackage{multirow}
\usepackage{orcidlink}
\usepackage{babel}

\usepackage{parskip}

\begin{document}

\title{
Screening-controlled dynamical criticality in the quantum Hall regime}

\author{Tanima Chanda\orcidlink{0009-0003-3516-8890}}
\affiliation{Department of Physics, Indian Institute of Science, Bangalore 560012, India}

\author{Simrandeep Kaur\orcidlink{0000-0002-1460-0686}}
\affiliation{Department of Physics, Indian Institute of Science, Bangalore 560012, India}

\author{Anantbir Virk}
\affiliation{Indian Institute of Science Education and Research, Pune 411008, India.}

\author{Kenji Watanabe}
\affiliation{Research Center for Electronic and Optical Materials, National Institute for Materials Science, 1-1 Namiki, Tsukuba 305-0044, Japan}

\author{Takashi Taniguchi}
\affiliation{Research Center for Materials Nanoarchitectonics, National Institute for Materials Science, 1-1 Namiki, Tsukuba 305-0044, Japan}

\author{G. J. Sreejith\orcidlink{0000-0002-2068-1670}}
\affiliation{Indian Institute of Science Education and Research, Pune 411008, India.}

\author{Yuval Gefen}
\affiliation{Department of Condensed Matter Physics, Weizmann Institute of Science, Rehovot 76100, Israel.}

\author{Aveek Bid\orcidlink{0000-0002-2378-7980}}
\email{aveek@iisc.ac.in}
\affiliation{Department of Physics, Indian Institute of Science, Bangalore 560012, India}

\begin{abstract}
At continuous electronic phase transitions, Coulomb interactions can modify the relation between length, energy, and temperature, but experimentally disentangling their effects on spatial versus dynamical criticality has remained difficult, since finite-temperature scaling alone measures only the combined exponent $\kappa = 1/(z\gamma)$. Here, we introduce two advances that resolve this limitation. First, by combining temperature scaling with independent current scaling, we separately extract the dynamical exponent $z$ and the localization-length exponent $\gamma$ at the quantum Hall plateau transition---rather than inferring one from an assumed value of the other. Second, using dual-graphite-gated graphene devices in which the effective Coulomb interaction range is tuned geometrically by the ratio of the magnetic length $l_B$ to the graphite-gate distance $d$, we track this separation across both screened and unscreened interaction regimes within the same device platform. Temperature scaling gives $\kappa \simeq 0.21$ in the screened regime and $\kappa \simeq 0.41$ in the unscreened regime; combining this with current scaling reveals that screening changes $z$ from $\simeq 1$ in the unscreened regime to $\simeq 2$ in the screened regime. In contrast, $\gamma$ remains close to $2.4$ throughout. Our results establish that gate-controlled screening selectively modifies the interaction-dependent dynamical sector of the quantum Hall transition, leaving the localization-length exponent $\gamma$ unchanged within experimental uncertainty. More broadly, this work establishes geometric screening as a versatile tool for controlling interactions and disentangling interaction and disorder effects in correlated two-dimensional systems, including fractional quantum Hall states, moiré materials, and other strongly localized electronic phases.
\end{abstract}
\maketitle

\paragraph{Introduction--}

In noninteracting systems, the critical properties of a continuous electronic phase transition are governed by diverging length and time scales, with the associated exponents encoding universal information about the transition~\cite{RevModPhys.69.315}. Coulomb interactions, however, also control the low-energy dynamics of real electronic systems and can therefore modify the relation between length, energy, and temperature. This raises a fundamental question: can screening alter the dynamical scaling while leaving the spatial critical exponent unchanged? More broadly, understanding how the range of the Coulomb interaction shapes criticality and localization is important for a wide class of strongly correlated two-dimensional systems, including fractional quantum Hall states and moir\'e materials, where interactions and disorder often play intertwined roles.

The integer quantum Hall (QH) plateau transition provides a well-controlled scenario to address this question~\cite{PhysRevLett.45.494}. In a perpendicular magnetic field $B$, disorder-broadened Landau levels contain localized electronic states separated by extended states near a critical energy $E_c$ close to the center of each Landau level~\cite{PhysRevLett.68.1375, PhysRevB.23.5632, PhysRevLett.54.831, PhysRevLett.61.593, Chalker1988, PhysRevLett.82.5100}. As the Fermi energy is swept through an extended state, the system undergoes a continuous localization-delocalization transition between adjacent QH plateaus~\cite{RevModPhys.69.315, RevModPhys.67.357, RevModPhys.80.1355}.
This plateau-to-plateau transition is parameterized by the Landau-level filling factor $\nu=nh/eB$ ($n$: areal carrier density, $h$: Planck's constant, and $e$: electronic charge), and occurs at a critical filling factor $\nu_c$. Near $\nu_c$, the localization length diverges as $\xi(\nu) \sim |\nu-\nu_c|^{-\gamma}$, while the characteristic time scale diverges as $\tau\sim\xi^z$~\cite{RevModPhys.67.357, PhysRev.177.952, PhysRevLett.102.216801, PhysRevLett.61.1297, Kaur2024, PhysRevLett.70.3796}. Here, $\gamma$ is the localization-length exponent and $z$ is the dynamical exponent which sets the finite-temperature cutoff of the critical divergence. For integer QH transitions, the commonly observed values $\gamma\approx2.4-2.6$, $z\simeq1$, and $\kappa=1/(z\gamma)\approx0.42$ have been discussed as signatures of universal critical behavior~\cite{PhysRevB.80.041304, PhysRevLett.61.1297, Kaur2024, PhysRevLett.128.116801, PhysRevLett.61.1297,RevModPhys.69.315,PhysRevLett.54.831,PhysRevLett.64.1437}.

\begin{figure*}[t]
\includegraphics[width=\textwidth]{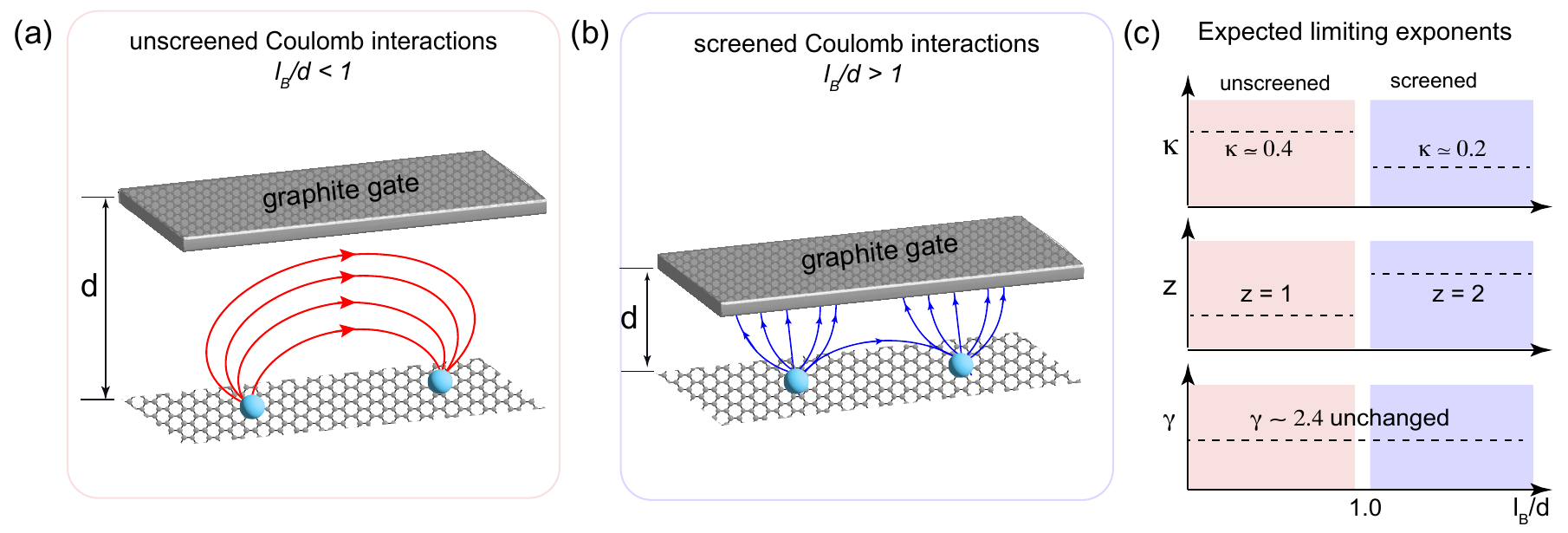}
\small\caption{\textbf{Screening-controlled crossover in quantum-critical scaling.}
Schematic of the interaction range in the unscreened regime, $l_B/d<1$ (a), and screened regime, $l_B/d>1$ (b). For $l_B/d<1$, the Coulomb interaction remains effectively long-ranged, whereas for $l_B/d>1$, the graphite gate screens the interaction between charge carriers in the graphene channel. (c) Expected limiting values of the finite-temperature scaling exponent $\kappa=1/(z\gamma)$, the dynamical scaling exponent $z$, and the localization-length exponent $\gamma$ across the screening-controlled crossover.
\label{fig:fig1}}
\end{figure*}

\begin{figure*}[t]
\includegraphics[width=\textwidth]{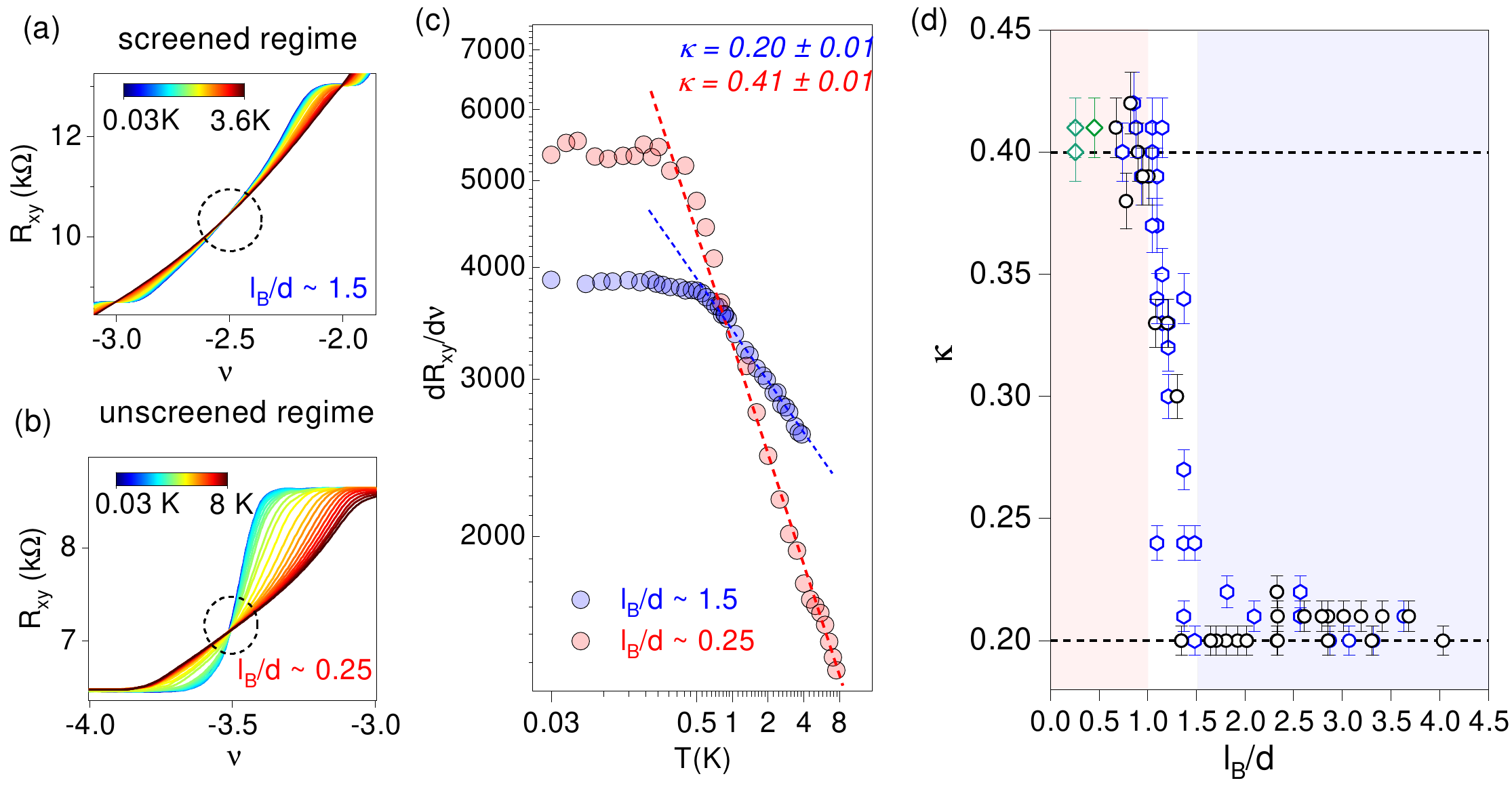}
\small\caption{\textbf{Temperature scaling of quantum-critical transport.} $R_{xy}$ as a function of filling factor $\nu$ across an integer QH plateau transition at representative temperatures in the (a) screened ($l_B/d\sim1.5$) and (b) unscreened ($l_B/d\sim0.25$) regimes. The common crossing of the isotherms (dashed outline) identifies the critical filling factor $\nu_c$. (c) Temperature dependence of the maximum Hall-resistance slope, $dR_{xy}/d\nu$, near $\nu_c$. Fits to $dR_{xy}/d\nu\propto T^{-\kappa}$ give $\kappa=0.20\pm0.01$ for $l_B/d\sim1.5$ and $\kappa=0.41\pm0.01$ for $l_B/d\sim0.25$. (d) Scaling exponent $\kappa$ as a function of $l_B/d$ measured in three devices (D1, blue open hexagons; D2, black open circles; D3, green open diamonds). Each point corresponds to an independently measured plateau transition. The data fall into two regimes: $\kappa\simeq0.4$ for $l_B/d\lesssim1$ (red shaded region) and $\kappa\simeq0.2$ for $l_B/d\gtrsim1$ (blue shaded region), separated by a narrow crossover. The error bars show fit uncertainty for each measurement; point-to-point scatter reflects device/filling variation, summarized in Table~\ref{tab:summary} of End Matter.
\label{fig:fig3}}
\end{figure*}

Electron-electron interactions add a further layer to this description, most notably in the quantum-critical regime near the plateau-to-plateau transition, where they determine the dynamical exponent $z$. Long-range Coulomb interactions are expected to give $z\simeq1$, whereas screened, effectively short-range interactions are predicted to shift the dynamical exponent to $z\simeq2$~\cite{PhysRevB.65.195316, PhysRevLett.76.4014, PhysRevResearch.4.033146, PhysRevB.61.8326, PhysRevLett.71.2638, PhysRevLett.82.5100} (Fig.~\ref{fig:fig1}). Work on quantum Hall criticality in GaAs- and InAs-based systems established the phenomenology of temperature and current scaling~\cite{PhysRevLett.61.1294, PhysRevB.50.14609, PhysRevLett.94.206807}. Building on this foundation, Tai~\textit{et al.}~\cite{Tai_2026} recently probed the effect of screening on $\kappa$ in a bilayer GaAs double-quantum-well system, providing the first indications that screening can modify quantum Hall criticality. However, their measurement accessed only the combined exponent $\kappa=1/(z\gamma)$, so $z$ could only be inferred by assuming $\gamma\simeq2.4$ rather than measured directly. Going beyond this, our work seeks a platform in which the screening length can be geometrically controlled and where the two critical exponents $z$ and $\gamma$ are \textit{independently} measured in the same device.

Here, we address this question using dual-graphite-gated graphene devices. The graphite screening gate is separated from the conducting charges in the graphene channel by an hBN spacer of thickness $d$, which defines the gate-screening length. Gate screening becomes effective when the relevant electronic length scale, set by the magnetic length $l_B=\sqrt{\hbar/eB}$, exceeds $d$.
Varying the magnetic field thus enables tuning the system from an unscreened regime ($l_B/d < 1$, Fig.~\ref{fig:fig1}(a)) to a screened regime ($l_B/d > 1$, Fig.~\ref{fig:fig1}(b)). Independently, adjusting the carrier density $n$ selects the filling factor $\nu$, allowing multiple plateau-to-plateau transitions to be probed at each value of $l_B/d$. This geometry makes it possible to investigate temperature scaling and current scaling of the same transitions within a single screening-controlled platform.

Using this device geometry, we show that controlling the interaction range has a strong effect on the critical exponents of the quantum-critical regimes. By combining temperature scaling, which yields $\kappa = 1/(z\gamma)$, with independent current scaling, which yields $z\kappa/(1+z)$, we determine $z$ and $\gamma$ independently and show that the evolution in critical scaling originates from the dynamical exponent. In contrast, the localization-length exponent remains essentially unchanged.
All measurements were performed on three devices with hBN spacer thicknesses of approximately 10 nm (D1), 11 nm (D2), and 25 nm (D3)~(Sec.~S1 of Supplemental Material (SM))~\footnote{The Supplemental Material includes details of device fabrication and data analysis.}.\\

\paragraph{Results:}
We first examine the effect of geometric screening on finite-temperature scaling at the plateau transition.
Figs.~\ref{fig:fig3}(a,b) show representative Hall-resistance isotherms across adjacent integer QH transitions for two values of $l_B/d$. In each case, the common crossing of the isotherms identifies the critical filling factor $\nu_c$ (dashed ellipse). Near $\nu_c$, the maximum Hall-resistance slope follows $dR_{xy}/d\nu\propto T^{-\kappa}$, where $\kappa=1/(z\gamma)$. Linear fits to $\ln(dR_{xy}/d\nu)$ versus $\ln T$ [Fig.~\ref{fig:fig3}(c)] give $\kappa=0.20\pm0.01$ for $l_B/d\sim1.5$ and $\kappa=0.41\pm0.01$ for $l_B/d\sim0.25$ for these particular data sets. The low-temperature deviation from linearity reflects finite-size effects~\cite{PhysRevLett.102.216801, Kaur2024}. Two additional methods for extracting $\kappa$ from $R_{xy}$ and $R_{xx}$ are presented in Sec.~S2 of the SM~\cite{Note1} and give consistent results. Exponents were extracted only over temperature and current ranges where these independent methods gave mutually consistent power-law behavior (Sec. S4 of the SM~\cite{Note1}).

\begin{figure*}
    \centering
    \includegraphics[width=0.9\textwidth]{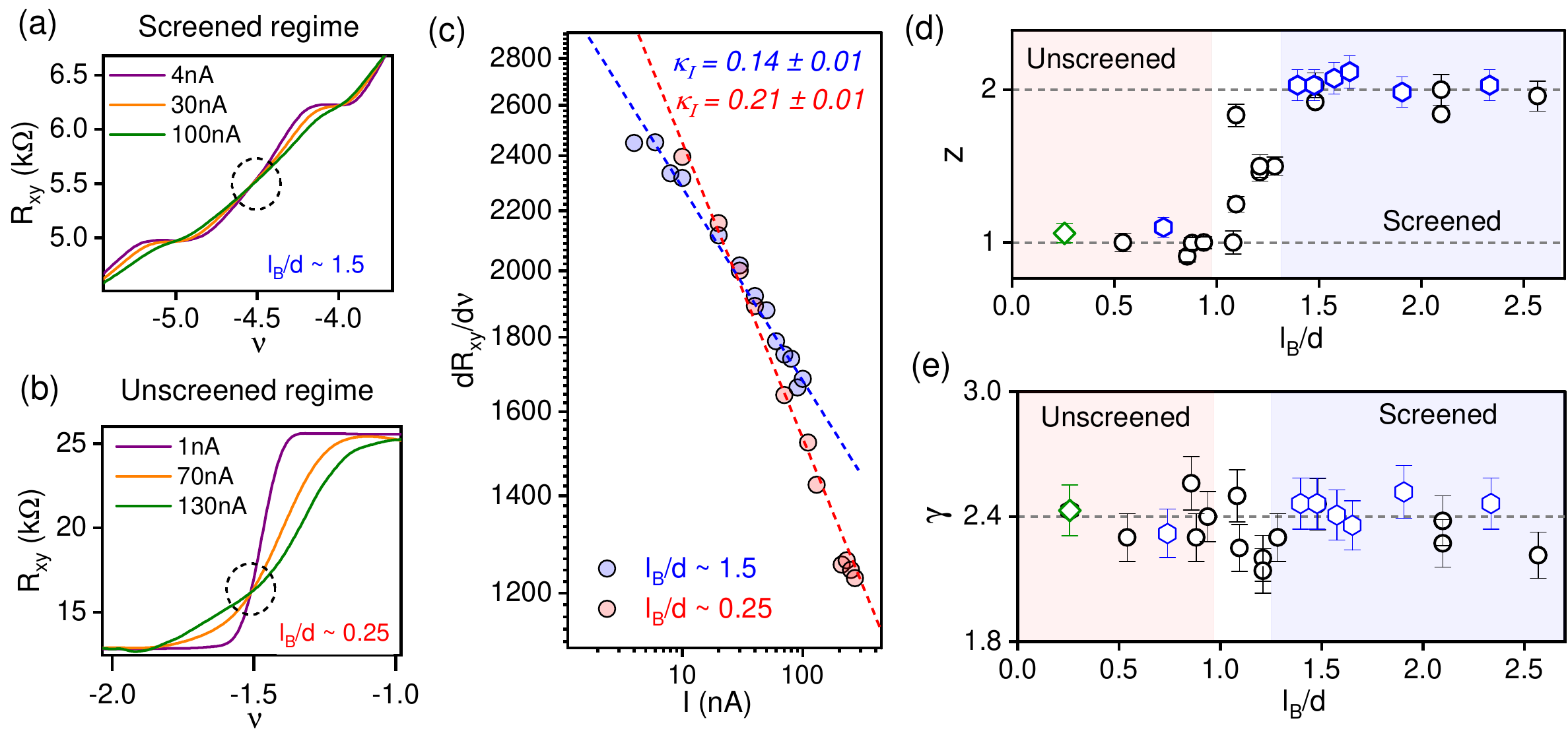}
\small\caption{\textbf{Current scaling separates the dynamical and spatial critical exponents.} $R_{xy}$ as a function of $\nu$ across an integer QH transition measured at different  $I$ in (a) the screened regime, $l_B/d\sim1.5$, and (b) the unscreened regime, $l_B/d\sim0.25$. The common crossing region identifies the critical filling factor $\nu_c$ (dashed circles). (c) $I$-dependence of the maximum of $dR_{xy}/d\nu$ near $\nu_c$. Fits to $dR_{xy}/d\nu\propto I^{-\kappa_I}$ yield $\kappa_I=z\kappa/(1+z)$. (d) Plots of $z$ versus $l_B/d$. Measurements from multiple integer QH transitions in three devices (D1, blue open hexagons; D2, black open circles; D3, green open diamonds) separate into two regimes: $z\simeq1$ for $l_B/d\lesssim1$, consistent with long-range Coulomb interactions, and $z\simeq2$ for $l_B/d\gtrsim1$, consistent with screened interactions. (e) Localization-length exponent $\gamma=1/(z\kappa)$ inferred from the independently extracted $z$ and $\kappa$, plotted versus $l_B/d$. $\gamma$ remains approximately constant near $2.4$ across the screened, crossover, and unscreened regimes. Error bars indicate the fit uncertainty of each individual measurement, while the scatter between points captures variation across devices and filling factors (Table~\ref{tab:summary} of the End Matter). \label{fig:fig4}}
\end{figure*}

Figure~\ref{fig:fig3}(d) summarizes $\kappa$ extracted from multiple plateau transitions across three devices. The error bars for each data point are for that particular measurement. Two regimes emerge: $\kappa\simeq0.41$ for $l_B/d\lesssim1$, where the interaction remains effectively long-ranged, and $\kappa\simeq0.21$ for $l_B/d\gtrsim1$, where the graphite gate screens the Coulomb interaction.  The ensemble-averaged values of $\kappa$ in these two regimes are tabulated in Table~\ref{tab:summary}.

\paragraph{Separating $z$ from $\gamma$ by current scaling:}
Since $\kappa=1/(z\gamma)$, the observed variation of $\kappa$ with $l_B/d$ raises an important question: does screening modify the localization-length exponent $\gamma$, the dynamical exponent $z$, or both? To distinguish between these possibilities, we determine $z$ independently from the current scaling of the plateau transition.
The origin of current scaling can be understood as follows~\cite{PhysRevB.50.14609}: At finite temperature, the divergence of the localization length near the critical filling factor, $\xi\sim|\nu-\nu_c|^{-\gamma}$, is cut off by the inelastic scattering length $L_{\mathrm{in}}$. Assuming $L_{\mathrm{in}}\propto T^{-p/2}$,  one can show that the effective electron temperature scales with an applied current $I$ as $T_e\propto I^{2/(2+p)}$. Substituting this into the temperature-scaling form gives $dR_{xy}/d\nu|_{\nu_c}\propto I^{-2\kappa/(2+p)}$. Using $pz = 2$, one obtains $dR_{xy}/d\nu |_{\nu_c}\propto I^{-z\kappa/(1+z)}$~\cite{PhysRevB.50.14609}. Thus, within the electron-heating scaling framework, current scaling provides an independent route to the dynamical exponent, provided $\kappa$ is known.

Figures~\ref{fig:fig4}(a, b) show representative $R_{xy}$ traces measured at different currents in the screened ($l_B/d\sim1.5$) and unscreened ($l_B/d\sim0.25$) regimes at $T=20$ mK.  As the current is reduced, the transitions sharpen and the maximum Hall-resistance slope, $dR_{xy}/d\nu$, increases. Linear fits of $\ln(dR_{xy}/d\nu)$ versus $\ln I$ [Fig.~\ref{fig:fig4}(c)] yield $\kappa_I=z\kappa/(1+z)$. The extracted exponents are consistent with prior studies where current scaling is governed by electron heating~\cite{PhysRevLett.89.276801}. Combining $\kappa_I$ with the independently measured $\kappa$ yields the dynamical exponent $z$, whose dependence on $l_B/d$ is summarized in Fig.~\ref{fig:fig4}(d) (two alternate methods used to extract $z$ from current scaling are discussed in Sec.~S3 of SM~\cite{Note1}). In the unscreened regime, $l_B/d<1$ (red shaded region), we find $z\simeq1$, consistent with a long-range Coulomb interaction setting the relevant dynamical energy scale. In contrast, in the screened regime, $l_B/d>1$, we find $z\simeq2$ (blue shaded region), as expected for screened, effectively short-range interactions.
The two ensemble-averaged values of $z$ are separated by $\sim 2.5\sigma$ (See \textit{End Matter}).
Combining the independently determined values of $z$ and $\kappa$ gives the localization-length exponent $\gamma=1/(z\kappa)$ [Fig.~\ref{fig:fig4}(e)], which remains close to the accepted integer-QH value $\gamma\simeq2.4$ throughout within experimental uncertainty (Table~\ref{tab:summary} of the End Matter).\\

\begin{figure}[t]
		\includegraphics[width=0.8\columnwidth]{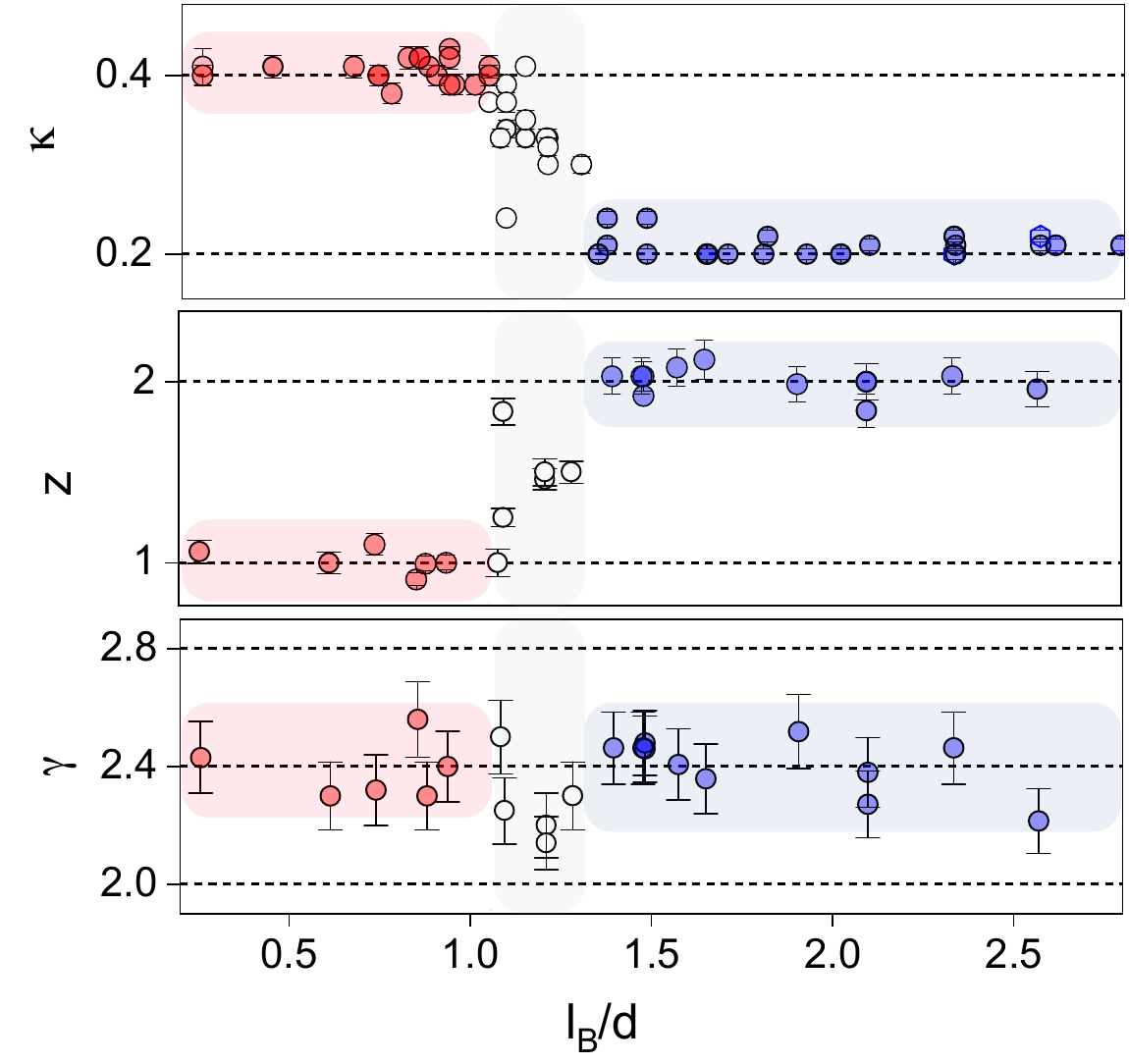}
\caption{\textbf{Evolution of the extracted exponents with $l_B/d$.}
Compilation of the experimentally extracted values of $\kappa$ and $z$ plotted versus $l_B/d$. Values of $\gamma$ are inferred from $\gamma=1/(z\kappa)$. The red, gray, and blue shaded regions indicate the unscreened, crossover, and screened regimes, respectively. Error bars show the fit uncertainty for each individual measurement; the point-to-point scatter within a regime reflects variation across devices and filling factors, summarized in Table~\ref{tab:summary} of End Matter.}
\label{fig:figS12}
	\end{figure}

\paragraph{Discussion:}

Gate-controlled screening selectively modifies the interaction-dependent sector of the quantum Hall transition. At the plateau transition, the localization length diverges in the thermodynamic limit, while finite temperature and finite sample size provide the experimental cutoffs. This divergent length controls the spatial critical scaling through $\xi\sim|\nu-\nu_c|^{-\gamma}$. The microscopic interaction range entering the Landau-level-projected dynamics, however, is governed by the projected Coulomb matrix elements. For a metallic screening gate a distance $d$ from the graphene channel, the interaction crosses over from long-ranged to screened at wave vector $q\sim1/d$. Since the Landau-level form factor weights the relevant interaction matrix elements at momenta of order $q\sim1/l_B$, the ratio $l_B/d$ provides a natural experimental parameter for tuning the effective interaction range. Thus, for $l_B/d\gtrsim1$ the gate substantially screens the interaction on the magnetic-length scale, whereas for $l_B/d\lesssim1$ the interaction remains effectively long-ranged.

Consistent with this, temperature scaling shows a crossover of $\kappa$ from $\simeq0.4$ in the unscreened regime to $\simeq0.2$ in the screened regime across multiple magnetic fields, filling factors, and three devices (see Sec.~S4 of the SM~\cite{Note1} for robustness checks). Current scaling shows that this change arises from a corresponding evolution of the dynamical exponent from $z\simeq1$ to $z\simeq2$, while the localization-length exponent remains close to $\gamma\simeq2.4$. Thus, screening changes the dynamical critical exponent while leaving the spatial localization exponent essentially unchanged (Fig.~\ref{fig:figS12} and Table~\ref{tab:summary} of End Matter).

The controlled crossover between screened and unscreened transport regimes becomes experimentally accessible in graphite-gated graphene because the hBN spacer fixes the distance between the conducting electrons and the metallic screening plane, while the magnetic field tunes the microscopic electronic length scale relative to this distance. This geometric control complements earlier GaAs studies of quantum Hall criticality and screening~\cite{PhysRevLett.61.1294, PhysRevB.50.14609, PhysRevLett.94.206807, Tai_2026}. More broadly, the screening-control strategy demonstrated here—using the gate distance $d$ as a geometric handle on the Coulomb interaction range—should be applicable to fractional QH states, correlated insulating phases in moir\'e systems, and other platforms where disentangling disorder and interaction effects remains a central challenge.

\section*{Data availability}
All data that support the plots within this paper and other findings of this study are available from the corresponding author upon reasonable request.

\section{Acknowledgments}
A.B. acknowledges funding from the Department of Science and Technology, Govt of India (SP/ANRF-24-0117). K.W. and T.T. acknowledge support from the JSPS KAKENHI (Grant Numbers 21H05233 and 23H02052) and World Premier International Research Center Initiative (WPI), MEXT, Japan.

\begin{center}
\textbf{\large End Matter}
\end{center}

\section*{Summary of critical exponents}

\begin{table*}[htbp]
\centering

\begin{tabular}{|l|c|l|l|l|}
\hline
\textbf{Exponent} & \textbf{Method of analysis} & \textbf{Screened} & \textbf{Unscreened} & $N$ \\
 & & $\mathbf{(l_B/d > 1)}$ & $\mathbf{(l_B/d < 1)}$ & \\ \hline
\multirow{5}{*}{$\kappa$} & $dR_{xy}/d\nu \propto T^{-\kappa}$ & $0.21\pm0.03$ & $0.41\pm0.06$ & $15$ \\ \cline{2-5}
 & $R_{xx}$ peak width & $0.20\pm0.05$ & $0.41\pm0.05$ & $15$ \\ \cline{2-5}
 & Finite-size scaling collapse & $0.21\pm0.04$ & $0.40\pm0.06$ & $5$ \\ \cline{2-5}
 & Scaling-error analysis & $0.21\pm0.04$ & $0.40\pm0.05$ & $5$ \\ \hline

\multirow{5}{*}{$z$} & $dR_{xy}/d\nu \propto I^{-\kappa_I}$ & $2.04\pm0.35$ & $1.02\pm0.20$ & $11$\\ \cline{2-5}
 & $R_{xx}$ peak width & $2.08\pm0.37$ & $1.04\pm0.21$ & $11$\\ \cline{2-5}
 & Finite-size scaling collapse & $2.00\pm0.38$ & -- & $5$ \\ \cline{2-5}
 & Scaling-error analysis & $2.05\pm0.36$ & -- & $5$\\ \hline

$\gamma$ & Estimated using $\kappa=1/(z \gamma)$ & $2.35\pm0.27$ & $2.4\pm0.28$ & $11$ \\ \hline
\end{tabular}
\caption{\textbf{Summary of critical exponents and scaling analysis.}
Values quoted are the mean $\pm$ standard error across $N$ independent plateau-to-plateau transition measurements per row, spanning three devices and multiple filling factors and $l_B/d$ values within each regime; $N$ is given in the rightmost column. Standard error is defined as
$\text{Standard Error} = \text{Standard Deviation}/\sqrt{N}$.
Individual fit uncertainties for representative datasets are typically
smaller than the ensemble spread reported here (see Fig.~\ref{fig:figS12} for single-dataset fit uncertainties); the values in this table instead capture variation across the full measurement ensemble. Based on the ensemble standard errors, the screened and unscreened values of $z$ are separated by $\sim\!2.5\sigma$.} \label{tab:summary}
\end{table*}

Table~\ref{tab:summary} summarizes the values of $\kappa$, $z$, and
$\gamma$ obtained using the independent methods described in this study and the SM~\cite{Note1}. Unlike the representative single-dataset fits reported there, the values quoted here are ensemble averages across all measured plateau transitions within each regime.

Note that the uncertainties quoted for representative single-dataset fits in both the main paper and the SM reflect fit precision for that measurement alone, not the ensemble spread reported here.

\section{Statistical significance of the change in $z$ across the screening-controlled crossover \label{sec:significance}}

To quantify the significance of the change in the dynamical exponent $z$
between the screened and unscreened regimes, we compare the
ensemble-averaged values reported in Table~~\ref{tab:summary}:
$z_{\text{unscreened}} = 1.02 \pm 0.20$ and
$z_{\text{screened}} = 2.04 \pm 0.35$, where the quoted uncertainties
are the standard errors of the mean across $N$ independent
plateau-transition measurements in each regime (see
 caption of Table~~\ref{tab:summary} for the definition of
the standard error).

Treating the two ensemble averages as independent estimates, the
uncertainty on their difference is obtained by adding the individual
standard errors in quadrature:
\begin{equation}
\sigma_{\text{diff}} = \sqrt{\sigma_{\text{screened}}^2 + \sigma_{\text{unscreened}}^2}
= \sqrt{(0.35)^2 + (0.20)^2} \approx 0.40.
\end{equation}
The significance of the separation is then
\begin{equation}
\frac{|z_{\text{screened}} - z_{\text{unscreened}}|}{\sigma_{\text{diff}}}
= \frac{|2.04-1.02|}{0.40} \approx 2.5\sigma.
\end{equation}

This calculation assumes that the screened- and unscreened-regime measurements are statistically independent, i.e., that no common systematic effect (such as a shared calibration offset or a correlated bias in the extraction of $\kappa$ or $\kappa_I$) shifts both ensemble averages in the same direction. We find no evidence for such a common systematic: the two regimes are measured using
different magnetic-field ranges, and the extraction procedures for $\kappa$ and $\kappa_I$ are applied
independently to each dataset (Secs.S2 and S3 of SM~\cite{Note1}).

We note that this significance is based on the Gaussian standard error
statistics and should be interpreted as indicative rather than exact,
given the finite number of measurements contributing to each
ensemble average ($N$ as tabulated in Table~~\ref{tab:summary}). The $\sim\ 2.5\sigma$ separation is consistent with the qualitative clustering of individual $z$ estimates near $z\simeq1$ and $z\simeq2$ visible in
Fig.~\ref{fig:fig4}(d)  and Fig.~\ref{fig:figS12}, where the tighter,
single-measurement fit uncertainties (typically $\pm0.05$--$0.06$) show comparatively little overlap between the two
regimes.

\clearpage

\addto{\captionsenglish}{\renewcommand{\refname}{Supplementary References}}

\renewcommand{\thesection}{S\arabic{section}}
\renewcommand{\thesubsection}{S\arabic{section}.\arabic{subsection}}
\renewcommand{\thefigure}{S\arabic{figure}}
\renewcommand{\thetable}{S\arabic{table}}

\setcounter{figure}{0}
\setcounter{equation}{0}
\setcounter{section}{0}
\setcounter{table}{0}

\begin{center}
    	{\textbf{\Large Supplemental Materials}}
\end{center}

{\hypersetup{linkcolor=black}
\renewcommand{\arraystretch}{1}

\section{Device fabrication}

ABA-stacked trilayer graphene (ABA-TLG), hBN, and graphite flakes were mechanically exfoliated onto $285$~nm SiO$_2$/Si substrates. ABA stacking was identified from optical contrast and Raman spectroscopy, using the characteristic splitting and intensity ratios of the Raman modes to distinguish ABA from ABC stacking~\cite{Cong2011, Nguyen2014}. The selected ABA-TLG flakes had a typical lateral size of $\sim20$--$30~\mu$m.

The van der Waals heterostructure was assembled using the standard dry pick-up and transfer technique~\cite{Dean2010, pizzocchero2016hot, PhysRevLett.129.186802, jat2024, Kaur2024, tsnc-4jjl}, yielding a final stack of graphite/hBN/ABA-TLG/hBN/graphite. This device structure enables dual-gated operation with atomically flat interfaces. One-dimensional edge contacts were defined by electron-beam lithography, followed by reactive ion etching in $\mathrm{CHF_3/O_2}$, and thermal evaporation of Cr/Pd/Au~\cite{doi:10.1126/science.1244358}. After this, the device was etched into a Hall bar geometry for longitudinal and Hall transport measurements. Typical field-effect mobilities extracted from low-field transport are listed in Table~\ref{tab:devices}. To prevent the formation of unintended $p$--$n$ junctions near the contacts, the graphene leads extending outside the dual-gated region were doped using the global SiO$_2$/Si back gate.

The carrier density $n$ and perpendicular displacement field $D$ were tuned independently via the top and bottom graphite gates, according to
\begin{equation}
D=\frac{C_{bg}V_{bg}-C_{tg}V_{tg}}{2\epsilon_0},
\qquad
n=\frac{C_{bg}V_{bg}+C_{tg}V_{tg}}{e},
\end{equation}
where $C_{bg}$ and $C_{tg}$ are the geometric capacitance per unit area of the back and top gates, respectively, and $V_{bg}$ and $V_{tg}$ are the corresponding gate voltages. The gate capacitance were extracted from the slopes of Landau fan diagrams in the quantum Hall regime.

We studied three devices in this work. The device details are provided in Table~\ref{tab:devices}.

\begin{table}
    \centering
    \begin{tabular}{|c|c|c|}
    \hline
        Device name & hBN thickness & mobility $\mathrm{cm^2V^{-1}s^{-1}}$\\ \hline
         D1& $10$~nm & $\sim 3.6\times 10^5$ \\ \hline
         D2& $11$~nm & $\sim 5.3\times 10^5$\\ \hline
         D3& $25$~nm & $\sim 3.1\times 10^5$\\ \hline
    \end{tabular}
    \caption{Details of the three devices measured in this study.}
    \label{tab:devices}
\end{table}

\section{Extraction of the temperature-scaling exponent $\kappa$ \label{sec:kappa}}

We estimate the values of $\kappa$ near criticality ($\nu \sim \nu_c$) using three independent methods: (i) analyzing the critical divergence of $d R_{xy}/d\nu$, (ii) probing the critical divergence of the inverse width of $R_{xx}(T)$, and (iii) a scaling analysis of $R_{xy}$ near the critical point.  In the main text, we discussed the data obtained using the first approach. We discuss the second and the third methods in the following two sections.

\subsection{Estimating $\kappa$ from the $T$-dependence of the width of $R_{xx}$}

We define $\Delta$ operationally as the separation between the two extrema in $dR_{xx}/d\nu$, corresponding to the full width of the transition region.
We analyze the temperature dependence of the transition width $\Delta$ as a function of $\nu$~\cite{PhysRevLett.61.1294}. The width follows the critical scaling, $\Delta\propto T^{\kappa}$~\cite{RevModPhys.67.357,ENGEL199013,Dodoo-Amoo_2014}. Fig.~\ref{fig:figS1}(a) shows representative plots of $R_{xx}$ versus $\nu$ (left axis, red lines) and $dR_{xx}/d\nu$ versus $\nu$ (right axis, blue line). The data were taken for  $l_B/d \approx 1.7$ ($B=2.2$~T). The slope of the linear fit to the logarithmic plot of $\Delta^{-1}$ versus $T$ yields $\kappa \approx 0.20 \pm 0.01$ (Fig.~\ref{fig:figS1}(b)). This value is consistent with that extracted from the $R_{xy}$ analysis.
For temperatures $T \lesssim 0.65$~K, $\Delta^{-1}$ saturates due to finite-size effects. In this regime, the transition width is no longer governed by critical scaling;  instead, it is limited by the finite sample size~\cite{PhysRevLett.102.216801, Kaur2024, PhysRevB.46.1596, PhysRevLett.67.883, PhysRevB.90.161408}.

\begin{figure*}[h]
	\includegraphics[width=\columnwidth]{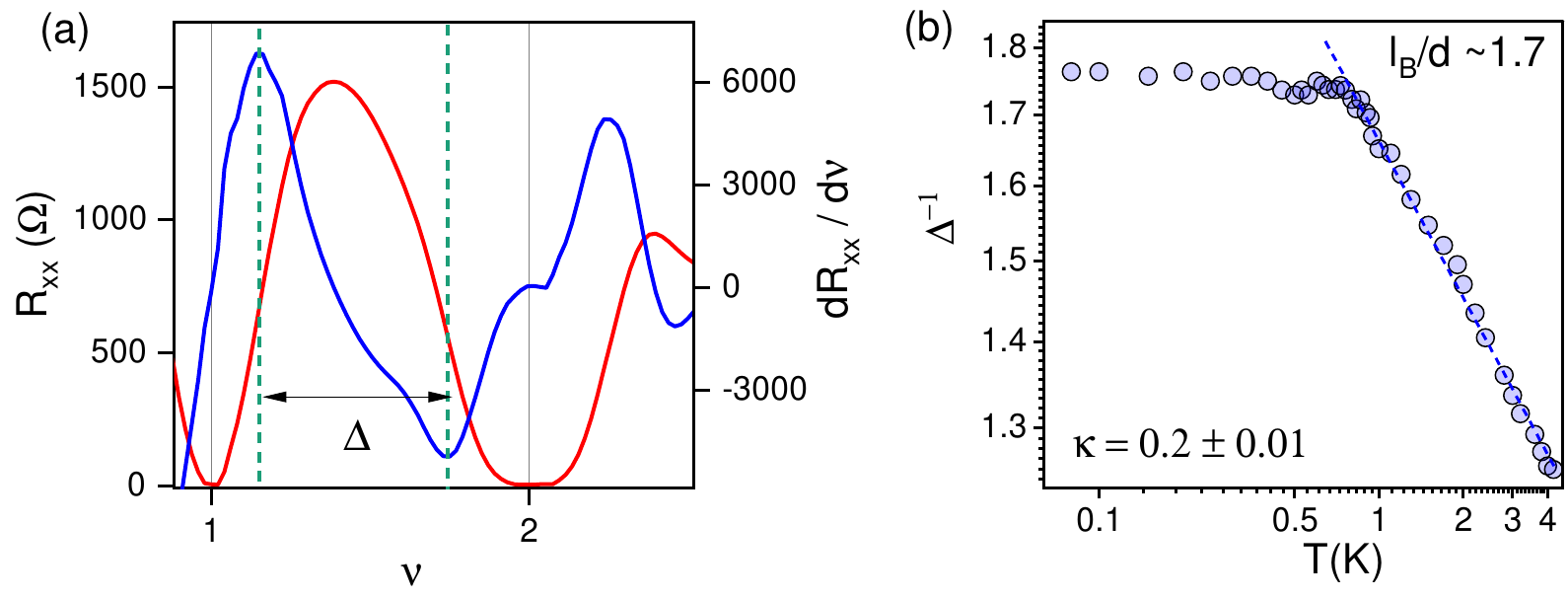}
	\small{\caption{\textbf{Estimating $\kappa$ from width of $R_{xx}$} (a)  Plots of $R_{xx}$ (left axis, red) and the corresponding $dR_{xx}/d\nu$ (right axis, blue) versus filling factor $\nu$ measured at $T=20$~mK. Green dashed lines mark the position of $\nu$ where $dR_{xx}/d\nu$ is maxima. $\Delta$ is the distance between two maxima. (b) Logarithmic plot of $\Delta^{-1}$ versus $T$. The black solid line is the linear fit to the data points.}
    \label{fig:figS1}}
\end{figure*}

\begin{figure*}[t]
	\includegraphics[width=\columnwidth]{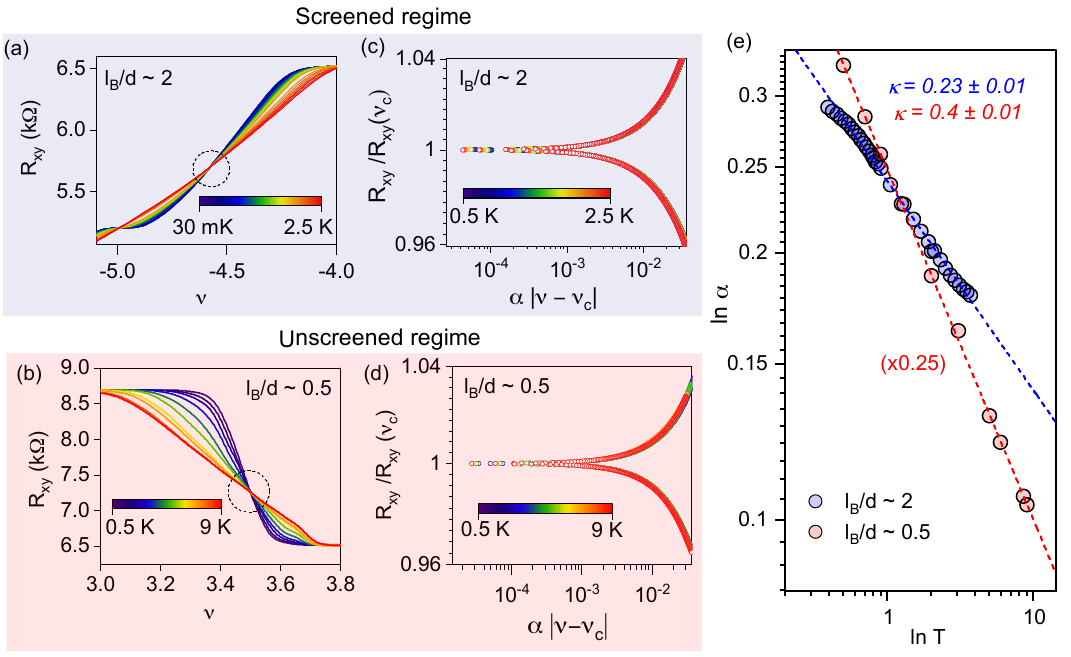}
	\small{\caption{\textbf{Finite-size scaling analysis.} Plot of $R_{xy}$ as a function of  $\nu$ for different temperatures measured for (a)  $l_B/d \sim 2$ and (b) $l_B/d \sim 0.5$. The dashed circle shows the $\nu_c$ for plateau-to-plateau transitions. (c-d) Finite-size scaling analysis near the critical point $\nu_c$. (e) Logarithmic plots of the scaling function $\alpha$ as a function of $T$. The blue (red) open circles are the data for $l_B/d \sim 2$ ($l_B/d \sim 0.5$), and the solid lines are the linear fits to the data points. The analysis of this particular data set yields  $\kappa = 0.23 \pm 0.01$ for the screened regime and $\kappa = 0.4 \pm 0.01$ for the unscreened regime. }
    \label{fig:figS2}}
     \end{figure*}

\subsection{Estimating $\kappa$ from finite-size scaling analysis}

To demonstrate the scaling properties of $R_{xy}$ close to $\nu_c$, we use~\cite{RevModPhys.67.357,PhysRevLett.61.1297,WEI199034}:
	\begin{equation}
		R_{xy}(\nu,T)=R_{xy}(\nu_c) f[\alpha(\nu-\nu_c)]
		\label{eq:fss1}
	\end{equation}
with $\alpha\propto T^{-\kappa}$, $f(0)=1$, and $f'(0)\neq 0$.
Figs.~\ref{fig:figS2}(a) and \ref{fig:figS2}(b) show representative $R_{xy}$ traces measured at $l_B/d\simeq2$ and $l_B/d\simeq0.5$, corresponding to the screened and unscreened regimes, respectively. The data have been replotted as $R_{xy}/R_{xy}(\nu_c)$ versus $\alpha|\nu-\nu_c|$ in Fig.~\ref{fig:figS2}(c) and Fig.~\ref{fig:figS2}(d). At every temperature, an optimal value of $\alpha(T)$ is chosen to collapse all the isotherms onto a single curve (the upper branch of the combined plot is for $\nu < \nu_c$, and the lower branch is for $\nu > \nu_c$). From the linear fits to the data points of $\alpha$ as a function of $T$ (Fig.~\ref{fig:figS2}(e)), we get $\kappa \sim 0.23 \pm 0.01$ for $l_B / d \sim 2$ and  $\kappa \sim 0.4 \pm 0.01$  for $l_B / d \sim 0.5$.

\subsection{Estimating $\kappa$ from scaling-error analysis}

As a third method to extract the value of $\kappa$ from the measured $R_{xy}(\nu, T)$, we perform an error analysis~\cite{Kaur2024}.  The procedure is illustrated in Fig.~\ref{fig:figS3} for $l_B/d \simeq2$ and in Fig.~\ref{fig:figS4} for $l_B/d\simeq~0.5$.
The data in each case are for two isotherms -- the red and blue curves correspond to $T=2.3$~K and $T=0.48$~K, respectively.
We start by treating $\kappa$ as a fitting parameter in Eqn.~\ref{eq:fss1}. Two sets of examples of this procedure are shown in Figs.~\ref{fig:figS3}(a--e) and Fig.~\ref{fig:figS4}(a--e). For each trial value of $\kappa$, the data are rescaled using Eqn.~\ref{eq:fss1}. The procedure is repeated for all temperatures.

For ideal scaling, all isotherm curves should collapse onto each other. Since the quality of collapse is difficult to judge visually, we quantify it by calculating the variance between the curves and use this variance as an error metric for the scaling collapse. The optimal value of $\kappa$ is then identified as the value that minimizes this error.

For the representative datasets shown here, this procedure gives $\kappa=0.23$ for $l_B/d\simeq2$ (Fig.~\ref{fig:figS3}(f)) and $\kappa=0.40$ for $l_B/d\simeq0.5$ (Fig.~\ref{fig:figS4}(f)).

\begin{figure*}[t]
		\includegraphics[width=0.9\columnwidth]{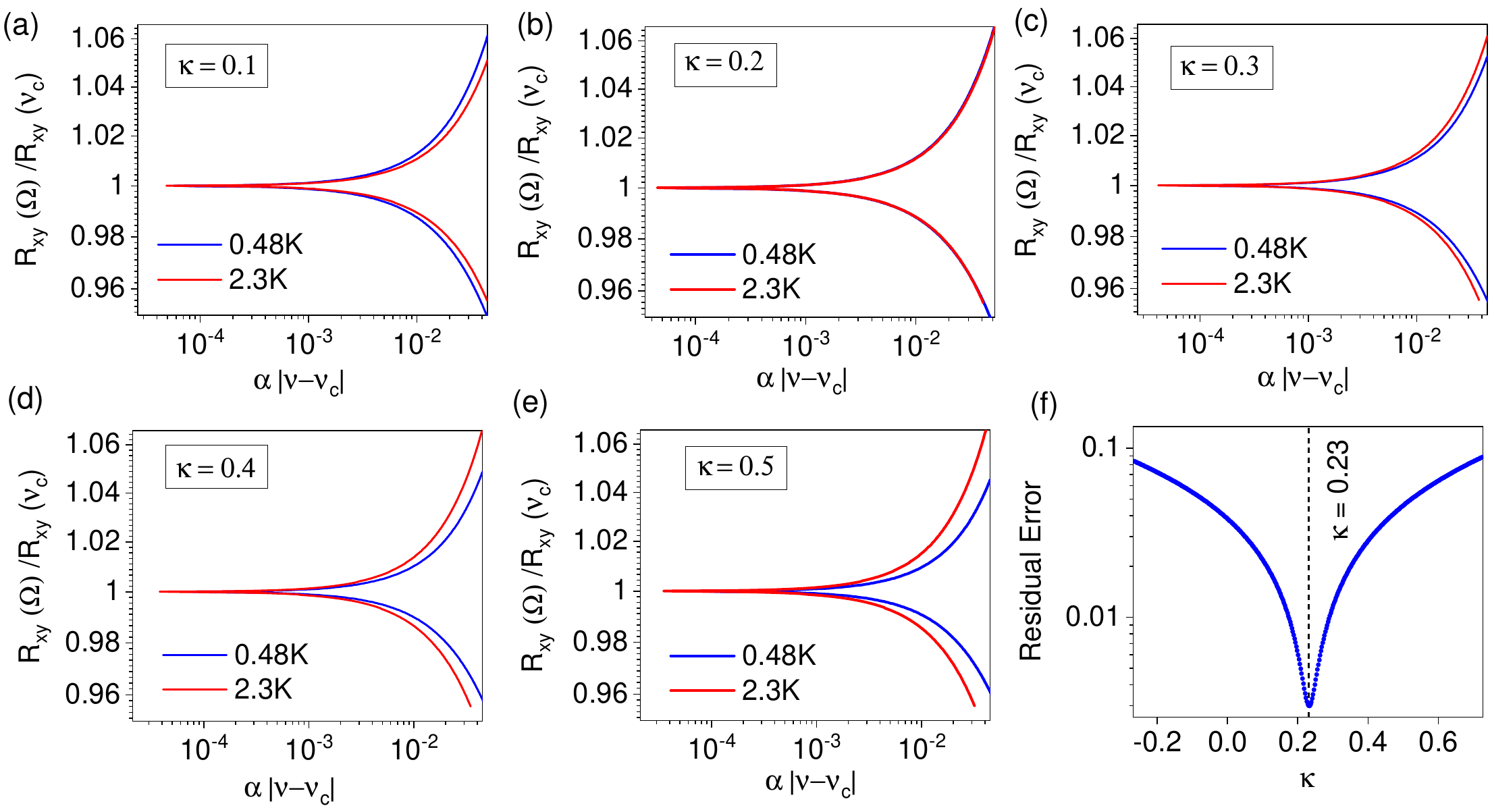}
		\small{\caption{\textbf{Estimating $\kappa$ from error analysis in screened regime $l_B/d\sim2$.} Finite-size scaling plot of $R_{xy}$ for  (a) $\kappa = 0.1$ (b) $\kappa = 0.2$ (c) $\kappa= 0.3$ (d) $\kappa= 0.4$ (e) $\kappa = 0.5$. (f) Plot of residual error in scaling near $\nu_c$ as a function of $\kappa$. The black dashed line marks the value of $\kappa$ where the residual error is minimum.}
\label{fig:figS3}}
	\end{figure*}
\begin{figure*}[h]
		\includegraphics[width=\columnwidth]{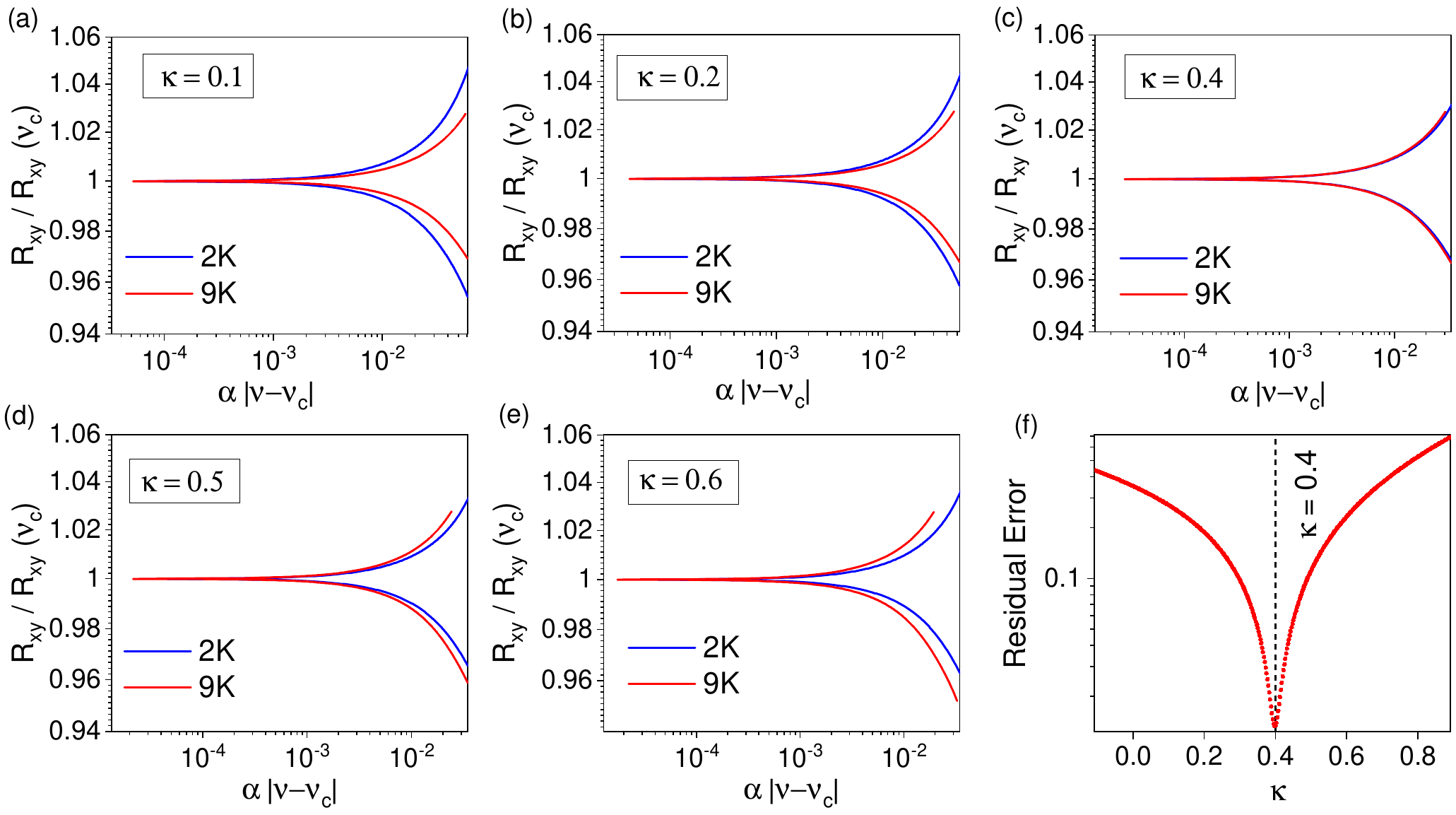}
    \small{\caption{\textbf{Estimating $\kappa$ from error analysis in unscreened regime $l_B/d\sim0.5$} Finite-size scaling plot of $R_{xy}$ for (a) $\kappa = 0.1$ (b) $\kappa = 0.2$ (c) $\kappa = 0.4$ (d) $\kappa = 0.5$ (e) $\kappa = 0.6$. (f) Plot of residual error in scaling near $\nu_c$ as a function of $\kappa$. The black dashed line marks the value of $\kappa$ where the residual error is minimum.}
\label{fig:figS4}}
	\end{figure*}

\clearpage
\section{Extraction of the dynamical exponent $z$ \label{sec:z}}
The dynamical exponent $z$ near the plateau transition $\nu \approx \nu_c$ is extracted using three methods: (i) the current dependence of $dR_{xy}/d\nu$, (ii) the current dependence of the inverse width of the $R_{xx}$ peak, and (iii) a scaling analysis of $R_{xy}$ near the critical point. The results from the first method are presented in the main text, while the latter two are discussed in the following sections.

\subsection{Estimating $z$ from $I$-dependence of the width of $R_{xx}$}
\begin{figure*}[h]
		\includegraphics[width=0.8\columnwidth]{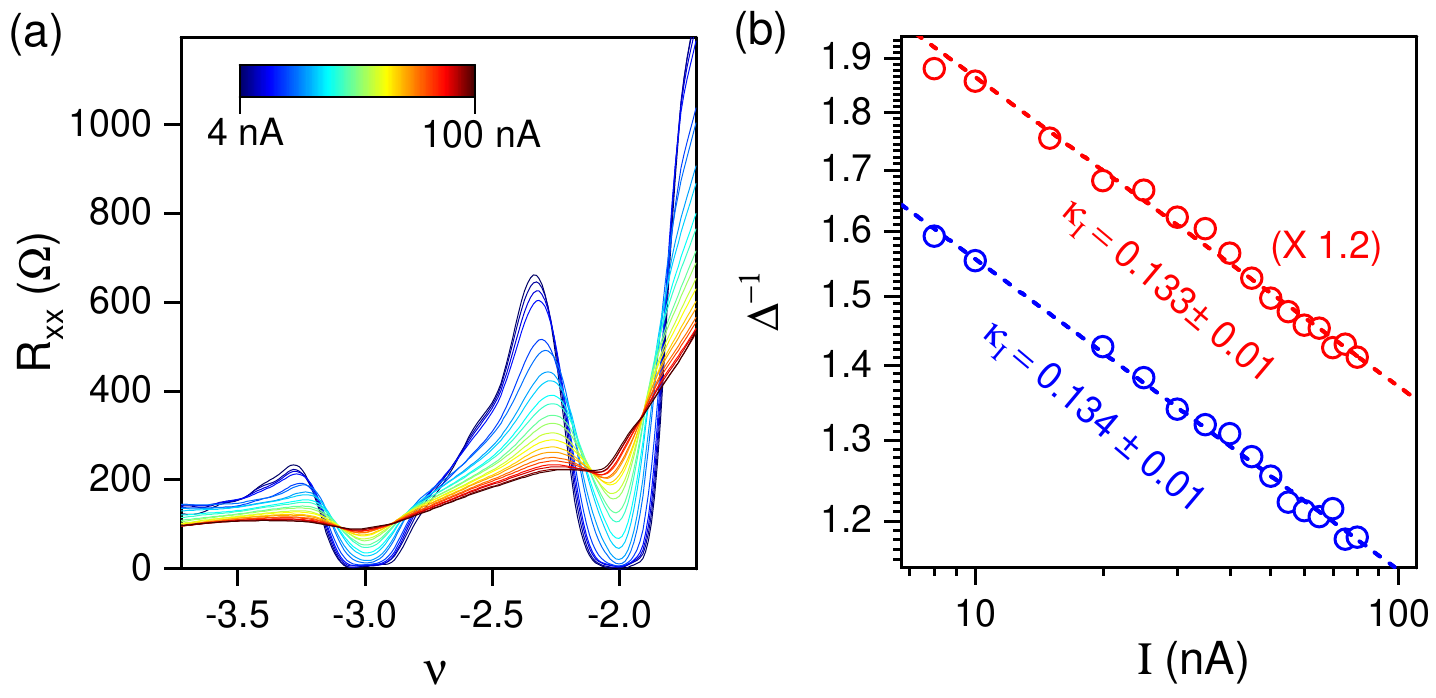}
		\small{\caption{\textbf{Estimation of dynamical exponent $z$ from width of $R_{xx}$} (a) Plot of longitudinal resistance $R_{xx}$ as a function of filling factor $\nu$ at different current bias measured at $B=3$~T and $T = 30$~mK. (b) Logarithmic plots of the inverse of the half-width of longitudinal resistance $R_{xx}$ versus $I$ for transitions between $\nu=-2$ to $\nu=-3$ and $\nu=-3$ to $\nu=-4$. Dashed lines are the linear fit to the data points.}
\label{fig:figS5}}
	\end{figure*}

The width $\Delta$ of the $R_{xx}$ peak is defined as the full width at half maximum as a function of the filling factor $\nu$. Representative plots of $R_{xx}$  measured at different bias currents at a fixed base temperature are shown in Fig.~\ref{fig:figS5}(a). As the current is increased, the transition broadens, and the inverse width decreases, following $\Delta^{-1}\propto I^{-\kappa_I}$ with $\kappa_I=z\kappa/(1+z)$~\cite{PhysRevB.50.14609}.
 A power-law fit to the data yields $\kappa_I = 0.134 \pm 0.01$ at $l_B/d \sim 1.5$  ($B=3$~T), as shown in Fig.~\ref{fig:figS5}(b). Combining this result with the temperature scaling exponent, $\kappa = 0.20 \pm 0.01$, obtained independently from temperature-dependent measurements, we estimate the dynamical critical exponent to be $z \approx 2.02 \pm 0.05$ for this representative dataset (see Table~I of End Matter in the main text for the ensemble-averaged value and its associated uncertainty).

\subsection{Estimating $z$ from finite-size scaling analysis}
As a second supplementary method to determine $z$, we perform a scaling analysis of the Hall resistance near the critical filling factor $\nu_c$. Close to the quantum Hall plateau transition, $R_{xy}$ is expected to obey the scaling relation~\cite{PhysRevLett.89.276801}
\begin{equation}
		R_{xy}(\nu,I)=R_{xy}(\nu_c) f[\alpha(\nu-\nu_c)]
		\label{eq:fss2}
	\end{equation}
with $\alpha\propto I^{-\kappa_I}$ and $\kappa_I=z\kappa/(1+z)$ ~\cite{PhysRevB.50.14609}; with $f(0)=1$, and $f'(0)\neq 0$.

  \begin{figure*}[t]
		\includegraphics[width=\columnwidth]{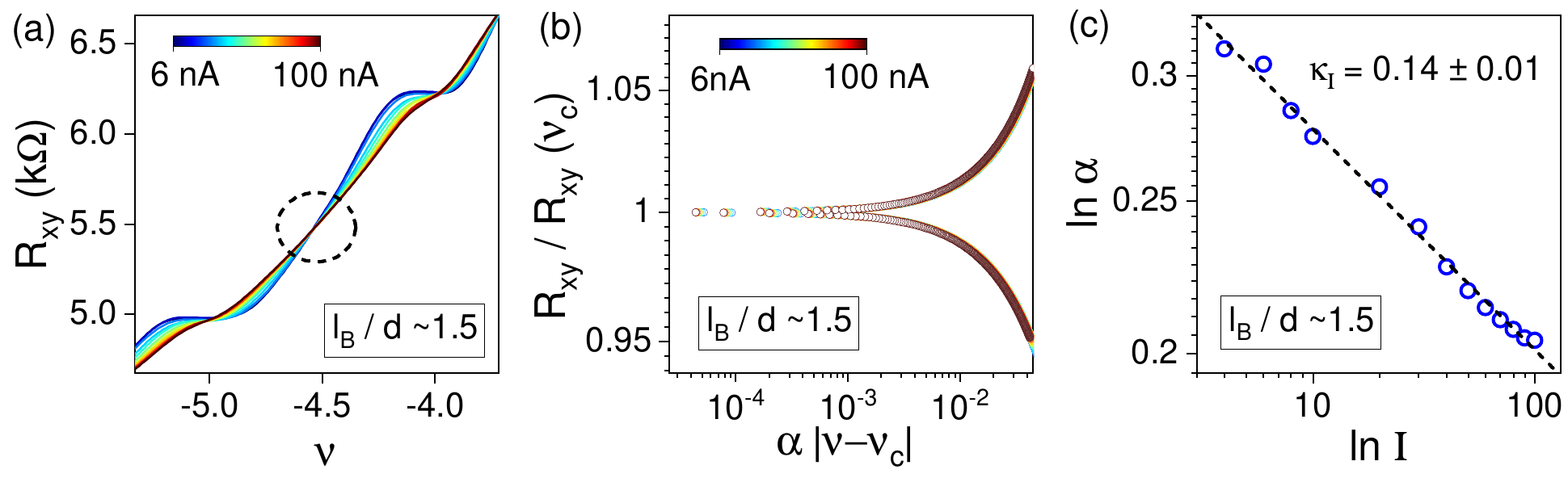}
		\small{\caption{\textbf{Estimating dynamical exponent $z$ from finite-size scaling analysis} (a) Plot of $R_{xy}$ as a function of filling factor $\nu$ for various bias currents ($6$~nA to $100$~nA). The dashed circle marks the critical filling factor $\nu_c$, where the $R_{xy}$ for different currents cross at a single point. (b) Finite-size scaling analysis of $R_{xy}$ at different bias currents near the critical point $\nu_c$. (c) The plot of the scaling function $\alpha$ as a function of $I$ on a log-log scale. The black dashed line is the linear fit to the data points.} \label{fig:figS6}}
	\end{figure*}

 $R_{xy}$ measured at different currents is shown in Fig.~\ref{fig:figS6}(a). To test the scaling behavior, the data are replotted as $R_{xy}/R_{xy}(\nu_c)$ versus $\alpha(\nu-\nu_c)$, as shown in Fig.~\ref{fig:figS6}(b). For each current, the scaling parameter $\alpha$ is adjusted to obtain the best collapse of all the $R_{xy}$ onto a single universal curve. The current dependence of the optimized scaling parameter follows a power law, $\alpha \propto I^{-\kappa_I}$ (Fig.~\ref{fig:figS6}~(c)). From the finite-size scaling analysis, we obtain $\kappa_I = 0.14 \pm 0.01$. Combining this result with the temperature scaling exponent $\kappa \simeq 0.21$ yields a dynamical critical exponent of $z = 2.0 \pm 0.06$ in the screened regime for this data set measured at $l_B/d \sim 1.5$ ($B=3$~T).

\subsection{Estimating $z$ from scaling-error analysis}
As an additional route to calculate the dynamical exponent $z$, we perform an error analysis using the measured $R_{xy}(\nu, I)$ data. For each trial value of $\kappa_I$, the data are rescaled and fitted to obtain the corresponding scaling curves as shown in Fig.~\ref{fig:figS7}(a--b). The quality of the scaling collapse is quantified by calculating the variance (least-squares residual) between the rescaled curves, which is taken as the fitting error. This error is then plotted as a function of the assumed $\kappa_I$. The optimum value of $\kappa_I$ is identified from the minimum of the error curve, corresponding to the best collapse of all the current-dependent $R_{xy}$ onto a single universal scaling function. In the screened regime where $l_B/d \approx 1.5$ ($B=3$~T), the error exhibits a well-defined minimum at $\kappa_I = 0.14$ as shown in Fig.~\ref{fig:figS7}(c). Combining this value with the independently obtained temperature scaling exponent $\kappa \simeq 0.21$ yields a dynamical critical exponent of $z \approx 2.0 \pm 0.06$, in good agreement with the values obtained from the other two independent scaling analyses.

 \begin{figure*}[t]
		\includegraphics[width=1\columnwidth]{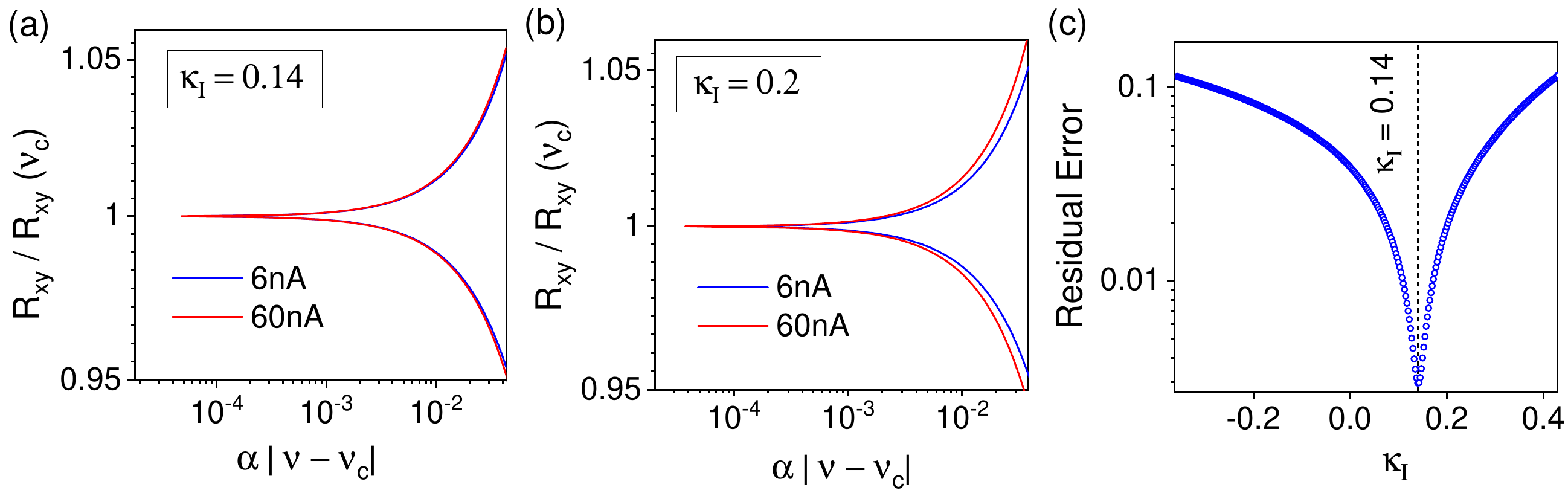}
		\small{\caption{\textbf{Calculating dynamical exponent $z$ from error analysis} Finite-size scaling plot of $R_{xy}$ for different values of $\kappa_I$ (a) $\kappa_I = 0.14$ (b) $\kappa_I = 0.2$. (c) Plot of estimated error in scaling as a function of $\kappa_I$.}
\label{fig:figS7}}
	\end{figure*}

\section{Finite critical window and robustness of exponent extraction}

In experiments on quantum Hall plateau transitions, critical scaling is accessible only over a finite temperature and filling-factor window. At the lowest temperatures, the divergence of the localization length can be cut off by finite sample size or finite inelastic length. At higher temperatures, thermal activation, Landau-level overlap, and disorder-dependent crossover scales can mask the critical power law. We therefore identify scaling windows in which independent procedures give consistent results.

For the temperature-scaling analysis, $\kappa$ is extracted from the maximum Hall-resistance slope $dR_{xy}/d\nu$ and independently from the full width at half maximum of the corresponding $R_{xx}$ peak. We also verify the stability of the extracted exponent against changes in the fitting window and across different plateau transitions, magnetic fields, and devices. The values of $\kappa$ quoted in the main text are extracted from the ranges over which these independent analyses are mutually consistent.

For the current-scaling analysis, $\kappa_I$ is extracted only over current ranges where $dR_{xy}/d\nu$ follows a stable power law and where excessive current-induced heating or low-current saturation is absent. The dynamical exponent $z$ and localization-length exponent $\gamma$ are then obtained only by combining values of $\kappa$ and $\kappa_I$ extracted within these mutually consistent scaling windows.

Thus, the quoted exponents are determined from temperature- and current-scaling analyses that remain stable against changes in fitting range, transition index, magnetic field, and device geometry.


%

\end{document}